\documentclass[]{aastex631}

\usepackage{gensymb}
\usepackage{amsmath}

\accepted{May 23, 2024}

\submitjournal{PSJ}

\shorttitle{Interpolation and synthesis of sparse samples in exoplanet atmospheric modeling}
\shortauthors{Haqq-Misra et al.}

\begin{document}

\title{Interpolation and synthesis of sparse samples in exoplanet atmospheric modeling}

\correspondingauthor{Jacob Haqq-Misra}
\email{jacob@bmsis.org}


\author[0000-0003-4346-2611]{Jacob Haqq-Misra}
\affiliation{Blue Marble Space Institute of Science, Seattle, WA, USA}
\affiliation{Consortium on Habitability and Atmospheres of M-dwarf Planets (CHAMPs)}

\author[0000-0002-7188-1648]{Eric T. Wolf}
\affiliation{Laboratory for Atmospheric and Space Physics, University of Colorado Boulder, Boulder, CO, USA}
\affiliation{Consortium on Habitability and Atmospheres of M-dwarf Planets (CHAMPs)}

\author[0000-0002-5967-9631]{Thomas J. Fauchez}
\affiliation{NASA Goddard Space Flight Center, Greenbelt, MD 20771, USA}
\affiliation{Integrated Space Science and Technology Institute, Department of Physics, American University, Washington DC}
\affiliation{Consortium on Habitability and Atmospheres of M-dwarf Planets (CHAMPs)}

\author[0000-0002-5893-2471]{Ravi K. Kopparapu}
\affiliation{NASA Goddard Space Flight Center, Greenbelt, MD 20771, USA}
\affiliation{Consortium on Habitability and Atmospheres of M-dwarf Planets (CHAMPs)}

\begin{abstract}
This paper highlights methods from geostatistics that are relevant to the interpretation, intercomparison, and synthesis of atmospheric model data, with a specific application to exoplanet {atmospheric} modeling. Climate models are increasingly used to study theoretical and observational properties of exoplanets, which include a hierarchy of models ranging from fast and idealized models to those that are slower but more comprehensive. Exploring large parameter spaces with computationally-expensive models can be accomplished with sparse sampling techniques, but analyzing such sparse samples can pose challenges for conventional interpolation functions. Ordinary kriging is a statistical method for describing the spatial distribution of a data set in terms of the variogram function, which can be used to interpolate sparse samples across any number of dimensions. Variograms themselves may also be useful diagnostic tools for describing the spatial distribution of model data in exoplanet {atmospheric} model intercomparison projects. Universal kriging is another method that can synthesize data calculated by models of different complexity, which can be used to combine sparse samples of data from slow models with larger samples of data from fast models. Ordinary and universal kriging can also provide a way to synthesize model predictions with sparse samples of exoplanet observations and may have other applications in exoplanet science.
\end{abstract}

\section{Introduction} \label{sec:intro}

The purpose of this paper is to demonstrate the application of methods from geostatistics to the analysis of exoplanet atmospheric models. Observations with JWST and other observatories are increasingly demonstrating feasibility for the spectral characterization of exoplanet atmospheres \citep[e.g.,][]{zieba2023no,madhusudhan2023carbon}. The need to understand the observational properties of known and plausible exoplanets has given rise to a diversity of climate models of different lineages and complexities that attempt to generalize the representation of climate in some way that extends to cases beyond Earth {\citep[see e.g,][for reviews]{shields2019climates,komacek2021constraining}}. 

The most complex of these are three-dimensional general circulation models (GCMs), which include explicit representation of atmospheric dynamics, radiative transfer, and surface exchange (sometimes with a dynamic ocean), as well as other parameterizations of physical processes {\citep[e.g.,][]{joshi2003climate,boutle2017exploring,way2017resolving,carone2018stratosphere,paradise2021exoplasim,wolf2022exocam,lefevre20213d}}. More idealized approaches include the use of energy balance models (EBMs), which calculate temperature across a one- or two-dimensional surface based on the balance of incoming starlight and outgoing infrared radiation and their effect on surface albedo {\citep[e.g.,][]{deitrick2018exo,kadoya2019outer,biasiotti2022eos,haqq2022energy,ramirez2024new}}. Even simpler models include 1-dimensional radiative convective equilibrium (RCE) models, which use a single vertical column to represent global planetary conditions {\citep[e.g.,][]{kasting1993habitable,hu2011radiative,kopparapu2013habitable,wordsworth2010gliese,windsor20211}}. This hierarchy of modeling approaches can provide the benefit of rapid exploration of parameter spaces with fast (but more idealized) models while also enabling more focused exploration of specific cases of interest with slow (but more comprehensive) models.

Validating exoplanet {atmospheric} models remains an ongoing challenge, due to the limited set of observations currently capable of constraining such model predictions. The model intercomparision projects of CUISINES (Climates Using Interactive Suites of Intercomparisons Nested for Exoplanet Studies, \url{https://nexss.info/cuisines/}) are developing standardized cases for comparing exoplanet {atmospheric} models in preparation for continued application to observations.  Most CUISINES recipes focus on comparing similar models for well-worn planet-atmosphere configurations.  However, one such exoplanet {atmospheric} model intercomparison project (exoMIP) is SAMOSA \citep{haqq2022sparse}, which will analyze and compare climate models of varying dimension and complexity across a broad sparse-gridded parameter space considering a hypothetical scenario of a terrestrial planet in a 15-day orbit around a 3000\,K blackbody star. The SAMOSA protocol defines a quasi-Monte Carlo sparse sample of points across a parameter space of surface pressure and incident instellation, which is designed to allow computationally-expensive GCMs to explore a broad range of the parameter space without becoming burdensome. The results of the SAMOSA exoMIP will include opportunities to compare different GCMs with each other as well as to compare GCM results with lower-dimensional models such as EBMs and RCE models that may be able to more completely explore the parameter space.  To yield the greatest benefit from the tapestry of simulations woven into SAMOSA, analysis methods are needed for interpolating climate data across the large sparse grided parameter space, and amongst the differing participant models.  The SAMOSA exoMIP is only one example of a broader need in exoplanet atmospheric modeling to compare GCM predictions and synthesize predictions across the modeling hierarchy.

Methods for analyzing sparse samples can be found in geostatistics, which were initially developed by Danie G. Krige and Herbert Sichel in the 1950s {(and later advanced by \citet{matheron1971theory})} to estimate ore reserves in South Africa. The general problem was that obtaining drill core samples is expensive, and so only a sparse number of samples could be obtained over the field of interest. Krige and Sichel developed methods for correlating and interpolating sparse samples that are today known as ``kriging'' and have been applied beyond mining to a wide range of problems, which include weather prediction and geographic information systems as well as climate science \citep[e.g.,][]{drignei2009kriging,garrigues2021capability} and astrophysics \citep[e.g.,][]{2017MNRAS.470.1121T}. The application of kriging to the analysis of climate models was suggested by \citet{drignei2009kriging}, who demonstrated that sparse samples of computationally-expensive GCM calculations can be combined with larger sets of less complex model calculations to make meaningful predictions without needing to run additional GCM cases. Exoplanet atmospheric modeling is well-suited to benefit from such methods, and this paper gives a brief presentation of how such methods may be useful in constraining exoplanet parameter spaces.

{The examples of kriging discussed in this paper are part of a broader class of ``emulation'' or ``surrogate modeling'' methods that are used in a range of disciplines and applications. A surrogate model or emulation is a simplified representation of a more complex or high-dimensional model, which can enable rapid retrieval of values across a large parameter space. While kriging gained traction within geostatistics communities, parallel developments of similar methods occurred with the advancement of computing simulations and the interest of the spatial statistics community \citep[][]{christianson2023traditional}, although today these disparate methods have converged for use by numerous communities. Many surrogate models will also leverage big datasets and large-scale computing in order to reduce the error threshold for predictions or meet other operational requirements \citep[e.g.,][]{alizadeh2020managing}, which could be based on machine learning techniques to construct the surrogate model. The development of surrogate models based on machine learning remains ongoing for a wide range of astronomical problems \citep[see e.g.,][for a review]{sen2022astronomical}.}

{The method most closely related to kriging is known as ``Gaussian process emulation,'' ``Gaussian process modeling,'' or ``Gaussian process regression,'' which has also been applied to many problems in astronomy \citep[e.g.,][]{lukemire2021statistical}. Gaussian process emulation is used as a synonym for kriging in some publications, and many of the differences between the two methods are only a choice in vocabulary (for example, the ``variogram'' in kriging is equivalent to the ``kernel function'' in Gaussian process emulation). But these two methods also differ in other important ways. Gaussian process emulation can include full automation for making inferences about patterns in datasets, whereas kriging requires human assessment as part of the algorithm. Gaussian process emulation may also be used to construct surrogate models that probabilistically emulate large data sets so that the resulting interpolations are accurate to a specified error threshold; by contrast, the kriging methods discussed in this paper provide exact interpolations at all data points. Both of these methods are distinct from machine learning, and in some cases they can outperform machine learning methods \citep[e.g.,][]{de2021introducing}, but Gaussian process emulation is better adapted than kriging for modeling very large data sets. The focus of this paper is on the analysis of sparse samples, so kriging is the method of choice, but it is worth noting that Gaussian process emulation could also be useful for certain problems in exoplanet atmospheric modeling.}

This paper proceeds by presenting the concept of the ``variogram'' (Section \ref{sec:variogram}), which represents the separation of pairs in the sample space. The variogram itself can serve as a diagnostic tool for assessing the distribution of data in a multi-dimensional space, which can be useful for analysis of a single model. Variograms also may be a useful aid in model intercomparison. The use of ``ordinary kriging'' as an interpolation tool for sparse samples is shown in Section \ref{sec:ordinary}, and the use of ``universal kriging'' to synthesize models of different complexity is shown in Section \ref{sec:universal}. The paper concludes in Section \ref{sec:future} by considering the possibility of incorporating observed quantities into such analyses as they eventually become available. The brief presentation of kriging in this paper is based on the text by \citet{wackernagel2003multivariate}. See this text and references therein for further and more comprehensive details regarding the derivation of kriging and its other potential applications.

\section{The Variogram and Model Intercomparison} \label{sec:variogram}

This section first describes the variogram and discusses its use in single-model analysis, followed by an example of using a variogram for an exoMIP. This approach calculates the functional form of a variable $z\left( \textbf{x} \right)$, where $\textbf{x}$ is a point in the spatial domain $\mathcal{D}$. In the first kriging problems, $z\left( \textbf{x} \right)$ represented the distribution of ore across a field of interest. The examples in this paper will use 
$z$
to represent temperature, with the spatial domain $\left( \mathcal{D} \right)$ as either a parameter space of surface pressure versus incident instellation or as a surface latitude-longitude map. In general, $z\left( \textbf{x} \right)$ and $\mathcal{D}$ can be any number of dimensions, although this paper will focus on a two-dimensional analysis.

\begin{figure}[ht!]
\centerline{\includegraphics[width=7.0in]{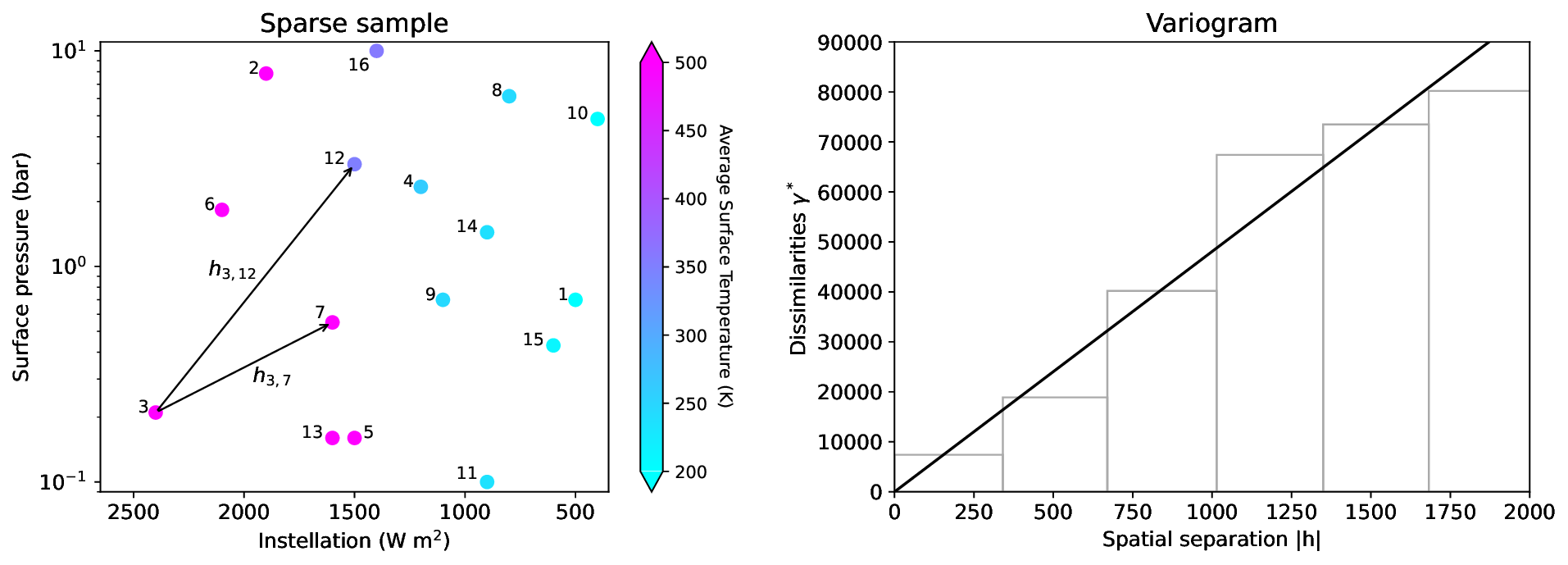}}
\caption{The left panel shows a uniformly-distributed sparse sample of points calculated by the ExoCAM model for the SAMOSA exoMIP \citep{haqq2022sparse}, with two examples of spatial separation vectors, $\textbf{h}$, drawn from point 3 at the base to points 7 and 12. The right panel shows the variogram from this sparse sample of ExoCAM calculations, which is calculated by assessing all pair of points in the sample, with gray bars showing the experimental variogram as a histogram of dissimilarities, $\gamma^{*}$, versus spatial separation, $|\textbf{h}|$, and a black line showing the best-fit theoretical variogram. The trend of increasing dissimilarity with increasing spatial separation is a typical pattern.
\label{fig:variogram}}
\end{figure}

Consider $z\left( \textbf{x}_{\alpha} \right)$ and $z\left( \textbf{x}_{\beta} \right)$ as two measured values in $\mathcal{D}$. Here, $\alpha$ and $\beta$ are used to respectively denote the first and second measured values, which in the context of this paper refers to two points calculated by a single climate model. The measure of ``dissimilarity'' between these two points is defined as $\gamma_{\alpha,\beta}^{*}=[z(\textbf{x}_{\alpha})-z(\textbf{x}_{\beta})]^{2}/2$. Defining a vector $\textbf{h}=\textbf{x}_{\alpha}-\textbf{x}_{\beta}$ to link these two points in space, the dissimilarity can be expressed more generally in terms of $\textbf{h}$ as
\begin{equation}
    \gamma^{*}(\textbf{h})=\frac{\left[ z(\textbf{x}_{\alpha}+\textbf{h})-z(\textbf{x}_{\alpha})\right]^{2}}{2}.
\end{equation}
Examples of the vector $\textbf{h}$ in the SAMOSA cases calculated with the ExoCAM GCM \citep{wolf2022exocam} are shown in Figure \ref{fig:variogram} (left). 

The plot of dissimilarities, $\gamma^{*}(\textbf{h})$ versus spatial separation, $\textbf{h}$, is known as the ``variogram cloud'' (not shown). The variogram cloud represents all pairs of cases in the sample but can be cumbersome to use directly, so an ``experimental variogram'' is typically constructed by binning the values of a variogram cloud into a histogram. An example of the experimental variogram for the ExoCAM SAMOSA cases is shown in Figure \ref{fig:variogram} (right, gray bars). The shape of the experimental variogram can further define a best-fit ``theoretical variogram,'' notated as $\gamma(\textbf{h})$, that can take several functional forms. The theoretical variogram for the ExoCAM SAMOSA cases is shown in Figure \ref{fig:variogram} (right, black line), which in this case is a linear function ($\gamma = s\cdot|\textbf{h}|+n$) with a slope of $s=48.1$ and a negligibly small value of the nugget term, $n$.

The variogram itself can provide an overview of the spread of data in a sample. The shape of the ExoCAM SAMOSA variogram (Fig. \ref{fig:variogram}, right) follows a typical variogram pattern of increased dissimilarity with increased spatial separation. A completely flat (horizontal) variogram would indicate no underlying spatial structure in the data. Abrupt changes in slope or a leveling-off of the variogram can provide information about the data variance, while discontinuities can signify a ``nugget effect'' in the data (analogous to gold nuggets hidden sporadically in a field). 

\begin{figure}[ht!]
\centerline{\includegraphics[width=7.0in]{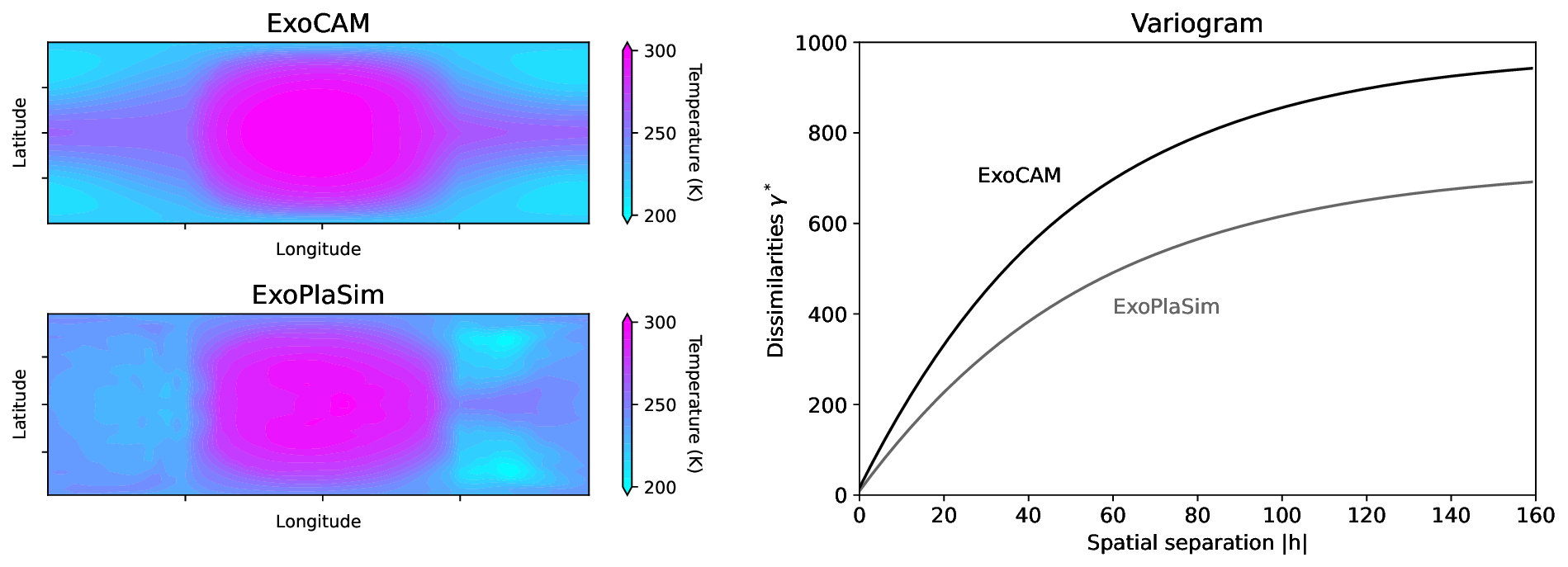}}
\caption{The left two panels show the distribution of surface average temperature for the ExoCAM (top) and ExoPlaSim (bottom) model calculations of SAMOSA case 4 (with 1200\,W\,m$^{-2}$ and 2.34\,bar). The right panel shows the theoretical variogram curves for both models. The warmer substellar point is one contributing factor to the ExoCAM variogram showing greater dissimilarities at all spatial separations when compared to ExoPlaSim.\label{fig:variogramcompare}}
\end{figure}

Variograms also can be instructive tools for conducting model intercomparison projects, especially within the exoplanet community. The example shown in Figure \ref{fig:variogram} illustrates the use of a variogram to analyze a set of points calculated by a single model across a defined parameter space, but variograms can in principle be applied to any set of points that are calculated across any spatial domain. For climate modeling, \citet{garrigues2021capability} demonstrated the capability for variograms to aid in comparing and quantifying surface spatial quantities calculated by different models. An example of this method is shown in Figure \ref{fig:variogramcompare}, which shows an intercomparison of two GCMs for the conditions of SAMOSA case 4 (with 1200\,W\,m$^{-2}$ and 2.34\,bar). The left panels show the average surface temperature calculated by the ExoCAM model (top) and the ExoPlaSim GCM of intermediate complexity \citep[bottom,][]{paradise2021exoplasim}, which show similar structure of a warm substellar point and cold night side as expected for a synchronously rotating planet. The surface temperature plots show ExoCAM with a warmer substellar point and much of the night side colder, when compared to ExoPlaSim. ExoCAM also features a narrow warm equatorial belt and greater axial symmetry, whereas ExoPlaSim has a more restricted warm region with asymmetric axial features. The variogram shown in the right panel of Figure \ref{fig:variogramcompare} provides a statistical representation of the surface temperature distribution for both models, with an exponential function ($\gamma = p\cdot(1-\exp{[-|\textbf{h}|/(r/3)]}+n$) providing the best fit. Here $p$ is the ``sill,'' equal to 978.9 for ExoCAM and 729.1 for ExoPlaSim, while $r$ is the ``range,'' equal to 144.7 for ExoCAM and 160.8 for ExoPlaSim. (The nugget term is negligible for both cases.) The sill and range provide diagnostic quantities in addition to the graphical representation of the variogram that can be utilized in model intercomparisons \citep[c.f.][]{garrigues2021capability}.

The variogram shows greater dissimilarities at all spatial separations for ExoCAM compared to ExoPlaSim, which is consistent with ExoCAM's warmer dayside and cooler nightside. Other models that calculate a surface temperature distribution for SAMOSA point 4 could also be included in such a variogram comparison, which could also be extended to any model variable beyond surface temperature.

This paper only seeks to demonstrate the potential for variograms to contribute to exoMIP analysis, and further examples will not be shown. But it should be emphasized that variograms can be used to assess the spread of model calculations across any spatial domain, with any number of dimensions. Figure \ref{fig:variogram} showed that variograms can be used for examining the behavior of a model across a defined parameter space. The two-dimensional parameter space covered by the SAMOSA cases are well-suited for calculating a variogram, and different GCMs that participate in SAMOSA will likely show different theoretical variograms. Figure \ref{fig:variogramcompare} showed that variograms can be used for comparing calculated model fields, and other exoMIPs could leverage the variogram as a tool for gaining deeper insight into model behavior. Significant divergence between the theoretical variograms of participating models could reveal important differences in the underlying model assumptions, while smaller differences in the slope of the theoretical variogram may also be instructive for understanding the differences in the spread of results for each model.

\section{Interpolation with Ordinary Kriging} \label{sec:ordinary}

The general kriging problem is to estimate a value at an arbitrary point $\textbf{x}_0$ in the spatial domain $\mathcal{D}$, given knowledge of a set of existing data values. If $n$ such data points are known, then a weighting process for calculating an arbitrary point in the domain takes the form
\begin{equation}
    Z^{*}(\textbf{x}_0)=\sum_{\alpha=1}^{n}w_{\alpha}Z(\textbf{x}_{\alpha}),\label{eq:Zstar}
\end{equation}
where $Z(\textbf{x}_{\alpha})$ is written as a capital letter to indicate that it is being treated as an intrinsic random function that depends on the sample points $\textbf{x}_{\alpha}$. The left-hand term $Z^{*}(\textbf{x}_0)$ denotes an intrinsic random function that can estimate a value at any location in the domain with the appropriate choice of weights, $w_{\alpha}$. By assuming that the underlying sampled data can be treated as a random function described by the variogram $\gamma(\textbf{h})$, the system of ordinary kriging can be written as
\begin{equation}
    \begin{cases}
        \sum_{\beta=1}^{n}w_{\beta}\gamma(\textbf{x}_\alpha-\textbf{x}_\beta)+\mu=\gamma(\textbf{x}_\alpha-\textbf{x}_0)\text{,    for }\alpha=1,...,n \\
        \sum_{\beta=1}^{n}w_{\beta}=1
    \end{cases}\,.\label{eq:ordinarykriging}
\end{equation}
where $\mu$ is the Lagrange parameter for solving this optimization problem. The optimal solution to the constraints in Eq. (\ref{eq:ordinarykriging}) provide weights for determining the functional form of $Z^{*}(\textbf{x}_0)$ in Eq. (\ref{eq:Zstar}).

\begin{figure}[ht!]
\centerline{\includegraphics[width=7.0in]{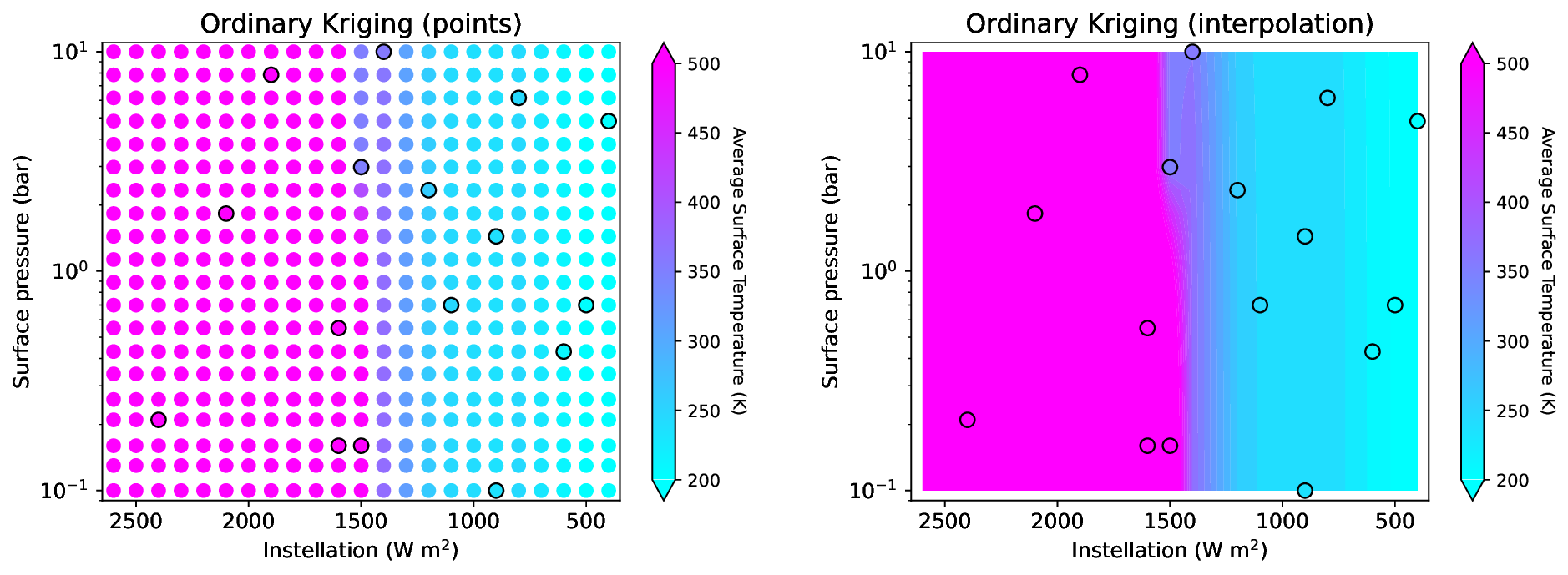}}
\caption{The values shown in the left panel are obtained by using ordinary kriging to map the sparse sample from Figure \ref{fig:variogram} onto a regularly-spaced grid, with the original sparse sample outlined in black. The right panel plots the kriged points with contour interpolation, with the original sparse sample shown as black dots.\label{fig:interpolation}}
\end{figure}

Sparse samples can be notoriously difficult to interpolate using conventional smoothing functions. Fortunately, ordinary kriging is an exact interpolator, such that $Z^{*}(\textbf{x}_0)=Z(\textbf{x}_0)$. Kriging can therefore be used to interpolate an entire spatial domain based on a sparse data sample by solving the ordinary kriging system Eq. (\ref{eq:ordinarykriging}) for a regular grid of points $\textbf{x}_0$ across $\mathcal{D}$. An example of this interpolation is shown for the ExoCAM SAMOSA cases in Figure \ref{fig:interpolation} (left), with data filled in over a regular grid based on the sparse sample of 16 points (outlined in black). This grid can then be interpolated through conventional contour smoothing function as shown in Figure \ref{fig:interpolation} (right). 

The interpolated solution obtained in Figure \ref{fig:interpolation} includes a region at large instellation where surface temperature exceeds 500\,K for all cases. These cases were identified as ``incipient runaway cases'' \citep{haqq2022sparse} that tend to show continued increases in pressure and temperature without easily reaching a final converged solution. For this kriging example, the incipient runaway cases were all fixed at a surface temperature value of 600\,K. The kriging interpolation finds little dependence on surface pressure at low instellation, with a transition region occurring near 1400\,W\,m$^{-2}$ from cooler to warmer climates. Points 12 and 16 (numbers indicated in Fig. \ref{fig:variogram}, left) contribute to the nonlinear behavior near this transition, with point 12 (at 1500\,W\,m$^{-2}$ and 2.98\,bar) at a surface temperature of 350.9\,K, cooler than the higher pressure point 16 (at 1400\,W\,m$^{-2}$ and 10.0\,bar). This feature is an example of how analysis with kriging can flag regions of interest in a sparsely sampled domain for further exploration by ``importance sampling'' in a subset of the domain.

\section{Intermodel Synthesis with Universal Kriging} \label{sec:universal}

The universal kriging approach considers situations in which sparse data samples reveal high-precision information at select locations in a domain but more readily available low-precision data is available everywhere in the domain. This is analogous to problems in geoexploration, with expensive drill core samples providing high-precision information at select locations and less expensive seismic surveys providing lower-precision information across the field of interest. For exoplanet {atmospheric} modeling, the high-precision information is obtained from a computationally-expensive GCM, while the lower-precision information across the domain is obtained with a more computationally-efficient model. 

With universal kriging, the random function is now written as
\begin{equation}
    Z(\textbf{x}) = m(\textbf{x}) + Y(\textbf{x}),\label{eq:Zuniv}
\end{equation}
where $m(\textbf{x})$ is known across the domain and referred to as the ``drift,'' and $Y(\textbf{x})$ is a second-order stationary random function with a covariance function $C(\textbf{h})$. The weighting function between $Z^{*}(\textbf{x})$ and $Z(\textbf{x})$ is the same as Eq (\ref{eq:Zstar}). The drift is written as a linear combination of deterministic functions, $f_l(\textbf{x})$, as
\begin{equation}
    m(\textbf{x})=\sum_{l=0}^{L}a_lf_l(\textbf{x}),\label{eq:muniv}
\end{equation}
where $L$ is the total number of drift data points, the $a_l$ values are non-zero coefficients, and $f_0(\textbf{x})=1$ by definition. The system of universal kriging can then be written as
\begin{equation}
    \begin{cases}
        \sum_{\beta=1}^{n}w_{\beta}C(\textbf{x}_\alpha-\textbf{x}_\beta)-\sum_{l=0}^{L}\mu_lf_l(\textbf{x}_\alpha)=C(\textbf{x}_\alpha-\textbf{x}_0)\text{,    for }\alpha=1,...,n \\
        \sum_{\beta=1}^{n}w_{\beta}f_l(\textbf{x}_\beta)=f_l(\textbf{x}_0)\text{,    for }l=0,...,L
    \end{cases}\,.\label{eq:univkriging}
\end{equation}
where $\mu_l$ are Lagrange parameters for solving this optimization problem. The optimal solution to the constraints in Eq. (\ref{eq:univkriging}) provide weights for determining the functional form of $Z^{*}(\textbf{x})$ using Eqs. (\ref{eq:Zstar}), (\ref{eq:Zuniv}), and (\ref{eq:muniv}). 

\begin{figure}[ht!]
\centerline{\includegraphics[width=7.0in]{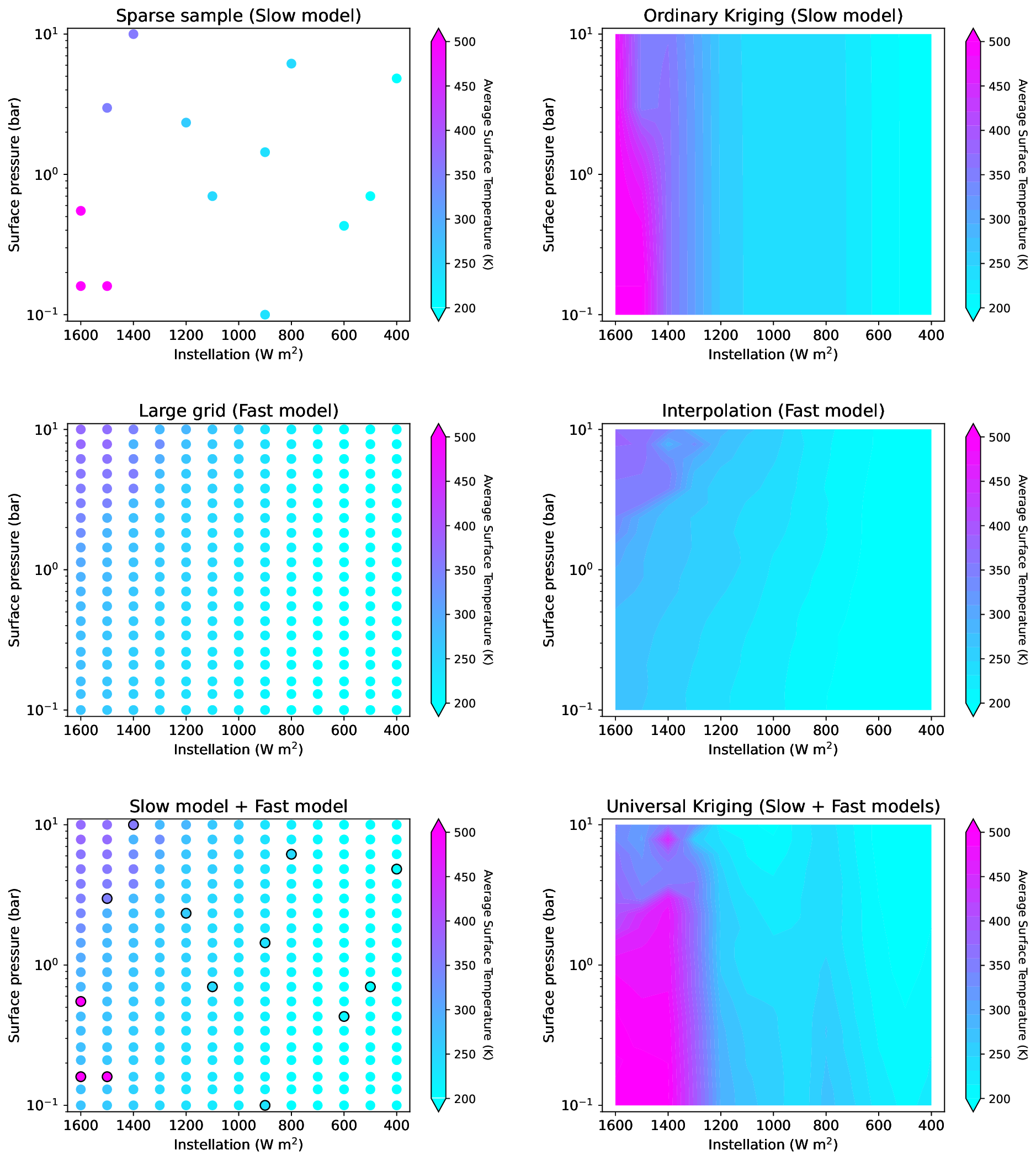}}
\caption{Kriging can be used to interpolate a sparse sample as well as to synthesize models of different complexity. The sparse sample calculated with the full-complexity ExoCAM model (top left) and ordinary kriging contour fit (top right) are identical to Figure \ref{fig:interpolation} but only using points with instellation 1600\,W\,m$^{-2}$ or less. A grid of solutions is shown using the intermediate-complexity ExoPlaSim model across the full parameter space (middle left), with the contour plot (middle right) indicating an under-prediction of warm cases compared to ExoCAM. The results shown in the bottom row are obtained by using universal kriging to map the ExoCAM sparse sample (bottom left, outlined points) across the parameter space by using the ExoPlaSim solutions (bottom left, colored points) as a drift function. Universal kriging can theoretically synthesize results from compatible models but may also introduce spurious results or nonphysical fine structure (bottom right).\label{fig:universal}}
\end{figure}

An example of universal kriging in the context of the SAMOSA cases is shown in Figure \ref{fig:universal}. For this example, the set of points has been limited to those with instellation 1600\,W\,m$^{-2}$ or less, and the surface temperature value of the incipient runaway cases has been set to 500\,K. The top left panel shows the ExoCAM sparse sample across the restricted parameter space, which is referred to as the ``slow model'' as it is ideal for in-depth exploration of a limited number of cases due to the computational expense of the model. The top right panel shows the result of interpolating these points with ordinary kriging, which is comparable to the results obtained over the larger parameter space in Figure. \ref{fig:interpolation}. 

The middle row of Figure \ref{fig:universal} shows calculations across the parameter space with the ExoPlaSim GCM of intermediate complexity, which makes simplifying physical assumptions that enable faster runtime than ExoCAM. This is referred to as the ``fast model'' because it can more efficiently explore the entire parameter space, albeit with reduced accuracy in the calculation of radiative transfer and other physical processes. The points in the middle left panel were obtained from previous calculations with ExoPlaSim \citep{paradise2021exoplasim}, and the middle right panel shows the resulting contour plot using conventional interpolation. There are some similarities and many differences between the ExoCAM (top right) and ExoPlaSim (middle right) surface temperature contour plots. The two models generally agree on the magnitude of cold temperatures for planets at high pressure, but the ExoPlaSim results show a stronger dependence on surface pressure. The incipient runaway regime is also absent in the ExoPlaSim results, and \citet{paradise2021exoplasim} acknowledged in the model description that ExoPlaSim is not ideally-suited for representing such moist and runaway atmospheres. These ExoPlaSim results thus are pushing the model to its realistic limits, which can be seen by the ``nugget'' that appears in the ExoPlaSim data at 1400\,W\,m$^{-2}$ and 7.85\,bar (which also appears in the original Fig. 25 by \citet{paradise2021exoplasim}). ExoPlaSim also tends to under-predict the warmest temperatures compared to ExoCAM in this domain.

The bottom row of Figure \ref{fig:universal} describes the approach of universal kriging by treating the ExoCAM cases as the high-precision data available at sparse locations and the ExoPlaSim cases as the drift function known across the domain. The left panel shows the superposition of the ExoCAM sparse sample (circled points) on top of the ExoPlaSim grid of points, which emphasizes the areas of agreement and disagreement noted in the preceeding paragraph. The right panel shows the result of universal kriging with these two models. The universal kriging interpolation shows magnitudes that match the predictions of ExoCAM but with enhanced contour structures that are informed by the ExoPlaSim results. Some of these features may indicate physical features that were not resolvable with the limited coverage of sparse sample, while other features may represent unphysical behavior that reveal key differences in the underlying model assumptions. The ``nugget'' point that appears in the universal kriging interpolation is one example that arises from limitations of ExoPlaSim, while the bimodal distribution of temperature for cold planets may also arise from non-physical reasons. 

Universal kriging provides a means of synthesizing results from complementary models, but such results should be utilized carefully. The results of universal kriging provide the best results when the sparse model and drift function have similar shapes. In many cases, models of different complexities may have non-linear differences that reveal unphysical behavior when universal kriging is attempted; in such cases, universal kriging may serve better as a tool for model intercomparison.

\section{Future Directions} \label{sec:future}

Ordinary kriging can play a role in fostering model intercomparisons across similar model types (i.e., comparing different GCMs). The variogram may be a useful tool for exoMIPs of all kinds for the analysis of variables in two- and three-dimensional space, which can even include the dimension of time. Variograms provide a statistical representation of the data distribution across the spatial domain (whether a geographical space or a parameter space), and a multi-model comparison of variograms could yield useful insight about important model behavior. For habitability problems, this could include outputs such as average albedo, surface habitability fraction, cloud coverage, and any other model quantities, in any number of dimensions.

Universal kriging may be useful for comparing or synthesizing results across different model types (i.e., comparing GCMs, EBMs, and RCE models). Specific problems could even utilize nested hierarchies of models; for example, this could take the form of first exploring a very large parameter space with a RCE model, followed by a more constrained but still comprehensive sampling of much of this domain with an EBM, and then concluding with a sparse set of GCM calculations in the regions of parameter space seemed to be the most physically interesting. If the shape of the solution is compatible among all three models, then this nested hierarchy could be combined through universal kriging to yield a prediction across the total parameter space but without requiring a prohibiting number of expensive GCM calculations. Such an approach may be particularly useful for exploring model behavior across a large number of parameters. The success of such an experimental approach will depend upon constructing models of different complexity that remain generally physically consistent with one another so as to be complementary and not introduce unphysical artifacts in the kriged interpolation.

{Kriging and Gaussian process emulation are powerful tools that can apply to a wide range of problems, but such methods also contain inherent limitations and uncertainties. The ordinary and universal kriging in this study perform exact interpolations at all data points, but this does not necessarily imply that the predictions between data points are correct. Gaussian process emulation likewise may relax the requirement for exact interpolations at data points, which tolerates some degree of uncertainty in the surrogate model. In both cases, Bayesian inference of this uncertainty can be done by performing additional model calculations (or collecting additional data) at unsampled points and assessing the prediction error. If these errors are too large, then additional kringing points are sampled (or additional points are included in the surrogate model) until an adequate prediction threshold is achieved. Such methods can be useful across specific operational domains, but Kriging and Gaussian process emulation could also introduce artifacts or nonphysical behavior that result from inhomogeneities in the underlying data. As research tools, emulation methods in general must be carefully evaluated against other ways of quantifying the data in order to ensure that any conclusions derived from emulations remain robust.}

The use of kriging to understand exoplanet {atmospheric} model results could also eventually incorporate exoplanet observations themselves as sparse data samples. The concept of the circumstellar habitable zone is one example of model-based prediction for exoplanets \citep{kasting1993habitable,kopparapu2013habitable}, and the eventual ability to observationally characterize the habitability of terrestrial planets will provide a basis for comparison with this theory \citep[e.g.,][]{lehmer2020carbonate,checlair2021theoretical}. In addition to direct comparison between model predictions and observations, a universal kriging synthesis of the two data sets could also be made by treating the exoplanet observations as the sparse data set and the modeling predictions as the drift function. Additional uses of kriging in this context and other applications in exoplanet science should continue to be explored.

\vspace{20pt}
The authors thank Nathan Mayne, Gavin Tabor, and Peter Challenor for helpful conversations. This material is based upon work performed as part of the CHAMPs (Consortium on Habitability and Atmospheres of M-dwarf Planets) team, supported by the National Aeronautics and Space Administration (NASA) under Grant No. 80NSSC21K0905 issued through the Interdisciplinary Consortia for Astrobiology Research (ICAR) program. ETW additionally acknowledges NASA Habitable Worlds Grant No. 80NSSC20K1421. The authors also acknowledge support from the Goddard Space Flight Center (GSFC) Sellers Exoplanet Environments Collaboration (SEEC), which is supported by the NASA Planetary Science Division's Research Program. 

\software{PyKrige \citep{pykrige}, ExoCAM \citep{wolf2022exocam}, ExoPlaSim \citep{paradise2021exoplasim}}

\bibliography{main}{}
\bibliographystyle{aasjournal}

\end{document}